\documentclass[twocolumn,secnumarabic,amssymb, amsmath, nofootinbib,tightenlines,
nobibnotes, aps, prl,epsfig]{revtex4}
\usepackage{graphicx}
\usepackage{dcolumn}
\usepackage{bm}
\begin{document}
\preprint{APS/123-QED}
\title{Approximation method for the nuclear structure functions at small $x$ }

\author{G.R.Boroun}%
 \email{ boroun@razi.ac.ir }
  \affiliation{Department of Physics, Razi University, Kermanshah
67149, Iran}%

\date{\today}
\begin{abstract}
We have shown that the relationship between the DIS structure
functions is stable for a nuclear target with mass number A at
$x{\leq}10^{-3}$. Numerical results are provided for the specific
nuclei 12C and 208Pb using the nuclear PDF parameterizations
implemented in the HIJING2.0 model. These results fall whithin the
EIC kinematic acceptance for heavy ion running. The ratio
$R^{A}_{F_{L}}$ is calculated as  the ratio
$R^{A}_{F_{2}}$ in the HIJING2.0 model.\\

\end{abstract}
 \pacs{***}
\maketitle

Some time ago, authors  proposed  approximation methods for the
proton structure functions to determine the gluon density at small
values of $x$ in Refs.[1,2]. In Ref.[1], the authors suggested a
connection between the longitudinal structure function at the
Bjorken variable $x$ and the gluon density at $2.5x$ using the
following formula for $n_{f}=4$ where $n_{f}$ is the number of
active flavors at the leading-order (LO) approximation
\begin{eqnarray}
F_{L}(x,Q^2){\simeq}\frac{10\alpha_{s}}{27\pi}xg(2.5x,Q^2).
\end{eqnarray}
Then, Prytz in Ref.[2] proposed a similar relationship between the
derivative of the proton structure function and the gluon density
at $2x$ as
\begin{eqnarray}
\frac{\partial{F_{2}(x,Q^2)}}{\partial{\ln}Q^2}{\simeq}\frac{10\alpha_{s}}{27\pi}xg(2x,Q^2).
\end{eqnarray}
These relationships were designed to isolate the gluon density
based on  expansions around $z=0$ and $\frac{1}{2}$ respectively.
Authors in Refs.[3] and [4] demonstrated that it is possible to
derive Eqs.(1) and (2) around arbitrary points $z=\alpha, \alpha'$
as they strongly depend on the momentum fraction carried by gluons
\begin{eqnarray}
F_{L}(x,Q^2){\simeq}\frac{10\alpha_{s}}{27\pi}xg(\frac{x}{1-\alpha}(\frac{3}{2}-\alpha),Q^2),
\end{eqnarray}
and
\begin{eqnarray}
\frac{\partial{F_{2}(x,Q^2)}}{\partial{\ln}Q^2}{\simeq}\frac{10\alpha_{s}}{27\pi}xg(\frac{x}{1-\alpha'}(\frac{3}{2}-\alpha'),Q^2).
\end{eqnarray}
The original versions of these approximation methods depend on the
transversal and longitudinal structure functions at small values
of $x$.\\
Nuclear structure functions, when compared to those for free
nucleons, are necessary for a detailed understanding of nuclear
shadowing in nuclear deep inelastic scattering at small values of
$x$. For a nuclear target with a mass number A, nuclear parton
distribution functions (nPDFs) scales with A. Therefore, the
nuclear structure functions can be expressed as
\begin{eqnarray}
F^{A}_{k=2,~L}(x,Q^{2})=<e^{2}>\sum_{i=s,g}\bigg{[}B_{k,a}(x){\otimes}xf^{A}_{i}(x,Q^{2})
\bigg{]}.
\end{eqnarray}
where the nuclear quark and gluon distributions are represented by
$xf^{A}_{s}(x,Q^{2})$ and $xf^{A}_{g}(x,Q^{2})$ respectively. Here
$<e^{2}>=\frac{\sum_{i=1}^{n_{f}} e_{i}^{2}}{N_{f}}$ denotes the
average charge squared, and the quantities $B_{k,i}(x)$ are the
known Wilson coefficient functions, with the parton densities
satisfying the renormalization group evolution
equations\footnote{The symbol $\otimes$ indicates convolution over
the variable $x$ by the usual form,
$f(x){\otimes}g(x)=\int_{x}^{1}
\frac{dz}{z}f(z,\alpha_{s})g(x/z)$.}.\\
The ratio of nuclear per nucleon density functions, known as small
$x$ shadowing [5], is defined as
\begin{eqnarray}
R_{i}^{A}=\frac{xf_{i}^{A}(x,Q^2)}{Axf_{i}^{p}(x,Q^2)},
\end{eqnarray}
where the nPDFs of a nucleus A with Z protons and N=A-Z neutrons
are given by
\begin{eqnarray}
f_{i}^{A}(x,Q^2)=\frac{Z}{A}f_{i}^{p/A}(x,Q^2)+\frac{N}{A}f_{i}^{n/A}(x,Q^2),
\end{eqnarray}
and the parton distribution functions for a bound proton and
neutron in a nucleus are denoted by $f_{i}^{p/A}(x,Q^2)$ and
$f_{i}^{n/A}(x,Q^2)$ respectively. Parametrization groups such as
K. J. Eskola, H. Paukkunen and C. A. Salgado (EPS) [6] and
HIJING2.0 [7,8] have
 discussed the ratio of nuclear structure functions, showing good
agreement with the ALICE experiment at LHC energies.\\
In this paper, we will focus on the possibility of directly
relating the gluon density to the nuclear longitudinal structure
function and the $Q^2$-dependence of the nuclear
structure function.\\
In perturbative Quantum Chromodynamics (pQCD), the nuclear
structure functions in the leading logarithmic approximation (LLA)
are as follows\footnote{At small $x$ ($x<10^{-2}$) the gluon
density is dominance and the sea-quark contribution can be
neglected. }
\begin{eqnarray}
\frac{\partial{F^{A}_{2}}(x,Q^2)}{\partial{\ln}Q^2}&{\simeq}&
\frac{5\alpha_{s}}{9\pi}\int_{0}^{1-x}dz
P_{qg}(1-z)G^{A}(\frac{x}{1-z},Q^2)\nonumber\\
&&=\frac{5A\alpha_{s}}{9\pi}\int_{0}^{1-x}dz
R^{A}_{G}(\frac{x}{1-z})P_{qg}(1-z)\nonumber\\
&&{\times}G(\frac{x}{1-z},Q^2),
\end{eqnarray}
and
\begin{eqnarray}
F^{A}_{L}(x,Q^2)&{\simeq}&
\frac{20\alpha_{s}}{9\pi}\int_{0}^{1-x}dz
C_{L,g}(1-z)G^{A}(\frac{x}{1-z},Q^2)\nonumber\\
&&=\frac{20A\alpha_{s}}{9\pi}\int_{0}^{1-x}dz
R^{A}_{G}(\frac{x}{1-z})C_{L,g}(1-z)\nonumber\\
&&{\times}G(\frac{x}{1-z},Q^2),
\end{eqnarray}
where $P_{qg}=z^2+(1-z)^2$, $C_{L,g}=z(1-z)$ and
$G(x,Q^2)=xg(x,Q^2)$.\\
The gluon distribution in Eqs.(8) and (9) can be expanded [3,4,9]
at arbitrary points $z=\alpha'$ and $\alpha$ due to the series
$\frac{x}{1-z}|_{z=\alpha,\alpha'}$ being convergent for
$|z-\alpha(\alpha')|<1$. Therefore Eqs.(8) and (9) become
\begin{eqnarray}
\frac{\partial{F^{A}_{2}}(x,Q^2)}{\partial{\ln}Q^2}&{=}&
\tau(A,x,Q^2)G\bigg{(}\frac{x}{1-\alpha'}\bigg{(}
1-\alpha'\nonumber\\
&&+\frac{\upsilon(A,x,Q^2)}{\tau(A,x,Q^2)}\bigg{)},Q^2\bigg{)},
\end{eqnarray}
and
\begin{eqnarray}
F^{A}_{L}(x,Q^2)&{=}&
\eta(A,x,Q^2)G\bigg{(}\frac{x}{1-\alpha}\bigg{(}
1-\alpha\nonumber\\
&&+\frac{\zeta(A,x,Q^2)}{\eta(A,x,Q^2)}\bigg{)},Q^2\bigg{)}.
\end{eqnarray}
The coefficients, in Eqs.(10) and (11), are defined as follows
\begin{eqnarray}
\tau(A,x,Q^2)&=&\frac{5A\alpha_{s}}{9\pi}\int_{0}^{1-x}({z^2+(1-z)^2})\nonumber\\
&&{\times}R^{A}_{G}\bigg{(}\frac{x}{1-z}\bigg{)}dz,\nonumber\\
\upsilon(A,x,Q^2)&=&\frac{5A\alpha_{s}}{9\pi}\int_{0}^{1-x}z({z^2+(1-z)^2})\nonumber\\
&&{\times}R^{A}_{G}\bigg{(}\frac{x}{1-z}\bigg{)}dz,\\
\eta(A,x,Q^2)&=&\frac{20A\alpha_{s}}{9\pi}\int_{0}^{1-x}{z}(1-z)R^{A}_{G}\bigg{(}\frac{x}{1-z}\bigg{)}dz,\nonumber\\
\zeta(A,x,Q^2)&=&\frac{20A\alpha_{s}}{9\pi}\int_{0}^{1-x}{z^2}(1-z)R^{A}_{G}\bigg{(}\frac{x}{1-z}\bigg{)}dz.\nonumber
\end{eqnarray}
According to the theory of parton shadowing, the reduction of the
gluon PDF should be proportional to the nucleon density times the
length of the nuclear matter at a given impact parameter.
Therefore, the authors in Refs.[7,8] have shown that the ratio
$R^{A}_{G}$ for light and heavy nuclei in the HIJING2.0 model
respectively are [7-10]
\begin{eqnarray}
R^{A}_{G}|_{\mathrm{light}}(z)&=&1+1.19(\ln{A})^{1/6}\bigg{(}z^3-1.2z^2+0.2z\bigg{)}\nonumber\\
&&-\frac{5}{3}{\times}0.22\bigg{(}A^{1/3}-1\bigg{)}^{0.6}\bigg{(}1-1.5z^{0.35}\bigg{)}\nonumber\\
&&{\times}\exp(-z^2/0.004)\bigg{]},
\end{eqnarray}
and
\begin{eqnarray}
R^{A}_{G}|_{\mathrm{heavy}}(z)&=&1+1.19(\ln{A})^{1/6}\bigg{(}z^3-1.2z^2\nonumber\\
&&+0.2z\bigg{)}-\frac{5}{3}{\times}0.22(1-(5~\mathrm{fm})^2/R_{A}^{2})\nonumber\\
&&{\times}\bigg{(}A^{1/3}-1\bigg{)}^{0.6}\bigg{(}1-1.5z^{0.35}\bigg{)}\nonumber\\
&&{\times}\exp(-z^2/0.004)\bigg{]},
\end{eqnarray}
where the nuclear radius is given by
$R_{A}=(1.12A^{1/3}-0.86A^{-1/3})~\mathrm{fm}$. Therefore, we find
\begin{eqnarray}
\frac{\partial{F^{A}_{2}}(x,Q^2)}{\partial{\ln}Q^2}/F^{A}_{L}(x,Q^2){=}
\frac{\tau(A,x,Q^2)}{\eta(A,x,Q^2)}{\times}\nonumber\\
\frac{G\bigg{(}\frac{x}{1-\alpha'}\bigg{(}
1-\alpha'+\frac{\upsilon(A,x,Q^2)}{\tau(A,x,Q^2)}\bigg{)},Q^2\bigg{)}}{G\bigg{(}\frac{x}{1-\alpha}\bigg{(}
1-\alpha+\frac{\zeta(A,x,Q^2)}{\eta(A,x,Q^2)}\bigg{)},Q^2\bigg{)}}.
\end{eqnarray}
Eq.(15) simplifies for the proton (with A=1) as defined in
Refs.[2,3] (with $\alpha'=\frac{1}{2}$) and Refs.[1,4] (with
$\alpha=0.666$) as
\begin{eqnarray}
\frac{\partial{F_{2}}(x,Q^2)}{\partial{\ln}Q^2}/F_{L}(x,Q^2){=}\frac{G(2x,Q^2)}{G(2.5x,Q^2)},
\end{eqnarray}
where it is simplified as the longitudinal structure function
scaled at $0.4x$ and  the derivative of $F_{2}(x,Q^2)$ with
respect to $\ln{Q^2}$ scaled at $0.5x$, by the following form [11]
\begin{eqnarray}
\frac{\partial{F_{2}}(\xi_{2}x,Q^2)}{\partial{\ln}Q^2}=F_{L}(\xi_{L}x,Q^2),
\end{eqnarray}
where $\xi_{2}=0.5$ and $\xi_{L}=0.4$.
\begin{figure}
\includegraphics[width=0.55\textwidth]{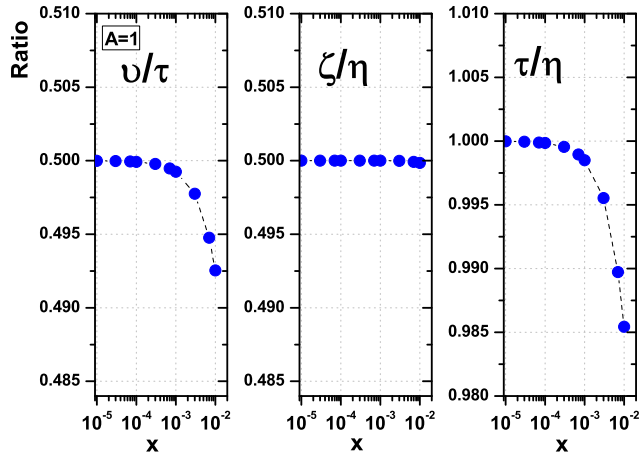}
\caption{The ratio of  $\upsilon/\tau$, $\zeta/\eta$ and
$\tau/\eta$ for the proton.}\label{Fig1}
\end{figure}
The ratio of coefficients (i.e., Eq.(12)), as
$x$ approaches 0, are chosen to be\\
 ${\upsilon}/{\tau}=0.5$,
${\zeta}/{\eta}=0.5$ and ${\tau}/{\eta}=1$.\\
 In Fig.1, we show
that these behaviors (i.e., ${\upsilon}/{\tau}=0.5{\pm}0.003$,
${\zeta}/{\eta}=0.5$ and ${\tau}/{\eta}=1{\pm}0.003$) in a wide
range of $x$ ($10^{-5}{\leq}x{\leq}10^{-2}$) are valid for
$x{\leq}10^{-3}$. Therefore, we conclude that the relation between
$F_{L}(x,Q^2)$ and
$\frac{\partial{F_{2}}(x,Q^2)}{\partial{\ln}Q^2}$  at the scales
$\xi_{L}$ and $\xi_{2}$ (i.e., Eq.(17)) is established at low $x$,
(i.e., $x{\leq}10^{-3}$).\\
 Now we apply this relationship owing
to Eq.(15) for light and heavy nuclei. In Figs.2 and 3, we observe
that the values of ratios (i.e., ${\upsilon}/{\tau}=0.5{\pm}0.02$,
${\zeta}/{\eta}=0.5{\pm}0.01$ and ${\tau}/{\eta}=1{\pm}0.02$) are
established for the light nucleus of C-12 and the heavy nucleus of
Pb-208 for $x{\leq}10^{-3}$ respectively. Therefore, the relation
between the structure functions (i.e., Eq.(17)) for light and
heavy nuclei is served by the following form at $x{\leq}10^{-3}$
as
\begin{eqnarray}
\frac{\partial{F^{A}_{2}}(\xi_{2}x,Q^2)}{\partial{\ln}Q^2}{\simeq}F^{A}_{L}(\xi_{L}x,Q^2).
\end{eqnarray}
This region (i.e., $x{\leq}10^{-3}$) will be probed in the
kinematical range of the future electron-Ion collider (EIC)
data\footnote{It is correct that the EIC$^{,}$s maximum
center-of-mass (COM) energy is roughly $140~ \mathrm{GeV}$ ($18~
\mathrm{GeV}$ electrons colliding with $275~ \mathrm{GeV}$ ions).
However, this is only possible for electron-proton collisions. The
maximum energy envisioned for electron-heavy ion runs would be
achieved by colliding $18~ \mathrm{GeV}$ electrons with $110~
\mathrm{GeV}$ ions for a $\sqrt{s}=89~ \mathrm{GeV}$} [12-14]. We
can rewrite Eq.(18) as
\begin{eqnarray}
F^{A}_{L}(x,Q^2){=}\frac{\partial{F^{A}_{2}}(\frac{\xi_{2}}{\xi_{L}}x,Q^2)}{\partial{\ln}Q^2}.
\end{eqnarray}
Now nuclear effects are applied to nuclear structure functions as
\begin{eqnarray}
R^{A}_{F_{L}}(x)F_{L}(x,Q^2)=R^{A}_{F_{2}}(\frac{\xi_{2}}{\xi_{L}}x)\frac{\partial{F_{2}}(\frac{\xi_{2}}{\xi_{L}}x,Q^2)}{\partial{\ln}Q^2}.
\end{eqnarray}
Usually, nuclear effects of $R^{A}_{F_{2}}(x)$ and $R^{A}_{G}(x)$
are parametrized in literature [15-23]. Here, we find that
\begin{eqnarray}
R^{A}_{F_{L}}(x)={\Phi}R^{A}_{F_{2}}(\frac{\xi_{2}}{\xi_{L}}x),
\end{eqnarray}
where
\begin{eqnarray}
\Phi=\frac{1}{F_{L}(x,Q^2)}\frac{\partial{F_{2}}(\frac{\xi_{2}}{\xi_{L}}x,Q^2)}{\partial{\ln}Q^2}.
\end{eqnarray}
The $\Phi$ function is dependent on the proton structure functions
as these functions are defined in literature\footnote{In Ref.[24],
the authors obtained the parametrization of the structure function
$F_{2}(x,Q^2)$ from a combined fit to HERA data in a wide range of
the kinematical variables $x$ and $Q^2$. Recently, in Ref.[25],
the authors described the longitudinal structure function
$F_{L}(x,Q^2)$ in momentum space using Laplace transform
techniques in a kinematical region of low values of the Bjorken
variable $x$. The authors in Refs.[26-33] defined the proton
structure functions in the color dipole picture (CDP), where the
photoabsorption cross sections at low $x$ are represented in terms
of the quark-antiquark proton cross sections by an ansatz for the
color-gauge-invariant interaction of the quark-antiquark  states
with the proton via two-gluon exchange.}. The value of $\Phi$ due
to the different methods [24-33] in a wide range of $x$ and $Q^2$
is $\Phi=1.0{\pm}0.2$. Therefore the nuclear effect
$R^{A}_{F_{L}}(x)$ is find
\begin{eqnarray}
R^{A}_{F_{L}}(x)=(1.0{\pm}0.2)R^{A}_{F_{2}}(\frac{\xi_{2}}{\xi_{L}}x),
\end{eqnarray}
where in the HIJING method this ratio is defined by the following
forms for light and heavy nuclei
\begin{eqnarray}
R^{A}_{F_{2}}|_{\mathrm{light}}(z)&=&1+1.19(\ln{A})^{1/6}\bigg{(}z^3-1.2z^2+0.2z\bigg{)}\nonumber\\
&&-\frac{5}{3}{\times}0.22\bigg{(}A^{1/3}-1\bigg{)}^{0.6}\bigg{(}1-3.5z^{0.5}\bigg{)}\nonumber\\
&&{\times}\exp(-z^2/0.01)\bigg{]},
\end{eqnarray}
and
\begin{eqnarray}
R^{A}_{F_{2}}|_{\mathrm{heavy}}(z)&=&1+1.19(\ln{A})^{1/6}\bigg{(}z^3-1.2z^2\nonumber\\
&&+0.2z\bigg{)}-\frac{5}{3}{\times}0.22(1-(5~\mathrm{fm})^2/R_{A}^{2})\nonumber\\
&&{\times}\bigg{(}A^{1/3}-1\bigg{)}^{0.6}\bigg{(}1-3.5z^{0.5}\bigg{)}\nonumber\\
&&{\times}\exp(-z^2/0.01)\bigg{]}.
\end{eqnarray}
In Fig.4, we plot $R^{A}_{F_{L}}$ in a wide range of $x$ for the
light nucleus of C-12 and the heavy nucleus of Pb-208.\\

In conclusion, our calculation of the DIS structure functions in a
nuclear target with mass number A at $x{\leq}10^{-3}$  utilizes
the DIS structure function relation for the proton at small $x$
with uncertainties less than $5\%$. We have compared our
relationship between the DIS nuclear structure functions in the
EIC domain. Additionally, the ratio $R^{A}_{F_{L}}$ derived from
$R^{A}_{F_{2}}$ as defined in the HIJING2.0 model.\\

\begin{figure}
\includegraphics[width=0.55\textwidth]{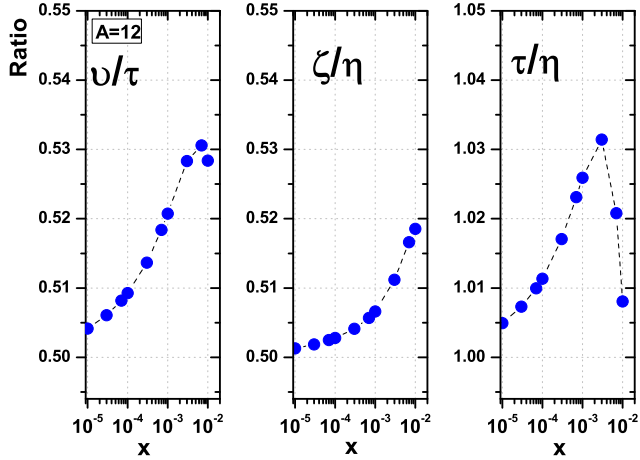}
\caption{The ratio of  $\upsilon/\tau$, $\zeta/\eta$ and
$\tau/\eta$ for the light nucleus of C-12.}\label{Fig2}
\end{figure}
\begin{figure}
\includegraphics[width=0.55\textwidth]{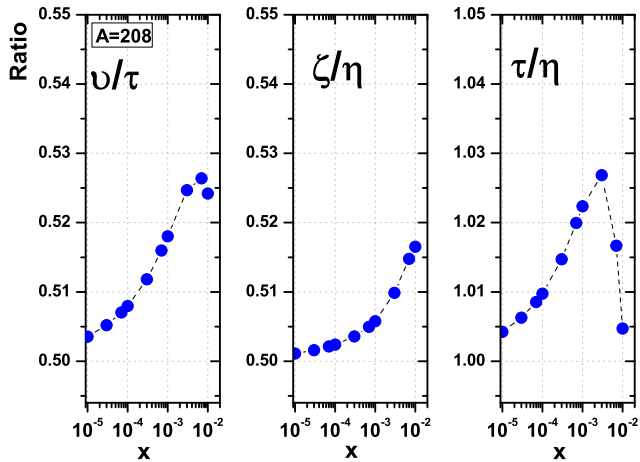}
\caption{The ratio of  $\upsilon/\tau$, $\zeta/\eta$ and
$\tau/\eta$ for the heavy nucleus of Pb-208.}\label{Fig3}
\end{figure}
\begin{figure}
\includegraphics[width=0.55\textwidth]{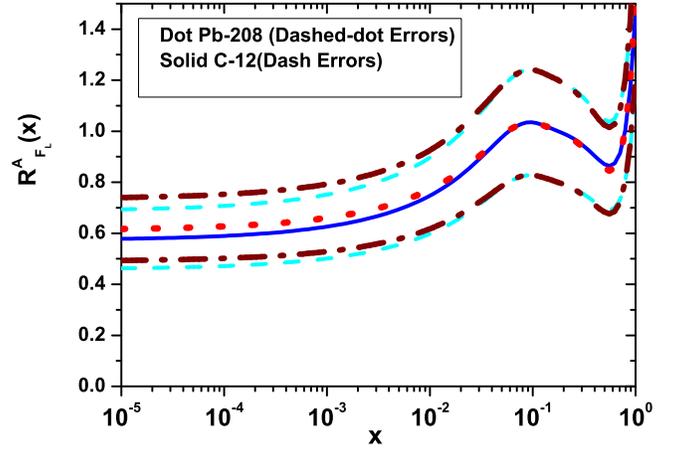}
\caption{$R_{F_{L}}$ for Pb (A=208) and C (A=12) as a function of
$x$. The uncertainties are proportional to the $\Phi$-function, as
shown in Eq.(22).}\label{Fig4}
\end{figure}

\subsection{ACKNOWLEDGMENTS}

G.R.Boroun would like to thank Professor V. P. Goncalves for
useful comments and invaluable support.\\


\section{References}
1. A. M. Cooper-Sarkar et al., Z. Phys. C: Part. Fields {\bf39},
281
(1988).\\
2. K. Prytz, Phys. Lett. B {\bf311}, 286 (1993).\\
3. M. B. Gay Ducati and P. B. Goncalves, Phys. Lett. B {\bf390},
401
(1997).\\
4. G. R. Boroun and B. Rezaei, Eur. Phys. J. C {\bf72}, 2221 (2012).\\
5. K.J. Eskola, H. Honkanen, V.J. Kolhinen and C.A. Salgado,
Phys.Lett.B {\bf532}, 222 (2002).\\
6. K. J. Eskola, H. Paukkunen and C. A. Salgado,
arXiv[hep-ph]:0802.0139.\\
7. S. -Y. Li and X. -N. Wang, Phys. Lett. B {\bf527}, 85 (2002);
Wei-tian Deng, Xin-Nian Wang and Rong Xu, Phys.Lett.B
{\bf701}, 133 (2011).\\
8. W. -T. Deng, X. -N.Wang and R. Xu, Phys.Rev.C {\bf83}, 014915
(2011); X.-N. Wang and M. Gyulassy, Phys. Rev. D 44,
3501 (1991); Comput. Phys. Commun. 83, 307 (1994).\\
9. J.Chen et al., arXiv [hep-ph]:2401.14904.\\
10. N.Armesto, Eur.Phys.J.C {\bf26}, 35 (2002).\\
11. G.R.Boroun, Phys.Rev.C {\bf97}, 015206 (2018).\\
12. A.Accardi et al., Eur.Phys.J.A {\bf52}, 268 (2016).\\
13. R. Abdul Khalek et al., (2021), arXiv:2103.05419
[physics.ins-det].\\
14. LHeC Collaboration and FCC-he Study Group, P. Agostini et al.,
J. Phys. G: Nucl. Part. Phys. {\bf48}, 110501 (2021).\\
15. K. J. Eskola, V. J. Kolhinen and C. A. Salgado, Eur. Phys. J.
C {\bf9}, 61 (1999).\\
16. D. de Florian and R. Sassot, Phys. Rev. D {\bf69}, 074028
(2004).\\
17. M. Hirai, S. Kumano and T. H. Nagai, Phys. Rev. C {\bf76},
065207 (2007).\\
18. N. Armesto, J. Phys. G {\bf32}, {R}367 (2006).\\
19. E.R. Cazaroto, F. Carvalho, V.P. Goncalves and F.S. Navarra,
Phys.Lett.B {\bf669}, 331 (2008).\\
20. [nCTEQ15 Collaboration] K.Kovarik et al., Phys. Rev. D
{\bf93},
085037 (2016).\\
21. [nNNPDF3.0 Collaboration] R.Abdul Khalek et al., Eur. Phys. J.
C {\bf82}, 507 (2022).\\
22. [EPPS21 Collaboration] K. J. Eskola, P. Paakkinen, H.
Paukkunen and C. A. Salgado,
Eur. Phys. J. C {\bf82}, 413 (2022).\\
23. V. P. Goncalves, EPJ Web of Conferences {\bf112}, 02006
(2016).\\
24.  M. M. Block, L. Durand, and P. Ha, Phys. Rev. D {\bf89},
094027 (2014).\\
25. G.R.Boroun and P. Ha, Phys. Rev. D {\bf109},
094037 (2024).\\
26. G. Cvetic, D. Schildknecht, B. Surrow and M. Tentyukov, Eur.
Phys. J. C {\bf20}, 77 (2001).\\
27. M. Kuroda and D. Schildknecht, Phys. Lett. B {\bf618}, 84
(2005); Phys. Lett. B {\bf670}, 129 (2008); Phys. Rev. D {\bf85},
094001 (2012); Int. J. Mod. Phys. A {\bf31}, 1650157 (2016).\\
28. G.R.Boroun, M. Kuroda and D. Schildknecht,
arXiv[hep-ph]:2206.05672.\\
29. D. Schildknecht and M. Tentyukov, arXiv[hep-ph]:0203028.\\
30. N. N. Nikolaev and B. G. Zakharov, Phys. Lett. B {\bf332}, 184
(1994).\\
31. N. N. Nikolaev and B. G. Zakharov, Z. Phys. C {\bf49}, 607
(1991).\\
32. N. N. Nikolaev and B. G. Zakharov, Z. Phys. C {\bf53}, 331
(1992).\\
33. J.J. Sakurai and D. Schildknecht, Phys. Lett.B {\bf40}, 121
(1972); B. Gorczyca and D. Schildknecht, Phys. Lett.B {\bf47}, 71
(1973); H. Fraas, B.J. Read and D. Schildknecht, Nucl. Phys. B
{\bf86}, 346 (1975); R. Devenish and D. Schildknecht, Phys. Rev. D
{\bf14}, 93 (1976).\\

\end{document}